\documentclass[preprint,aps,nofootinbib]{revtex4}
\usepackage{graphicx}
\usepackage{amsmath}
\usepackage{amsfonts}
\usepackage{amssymb}
\usepackage{color}%
\providecommand{\U}[1]{\protect\rule{.1in}{.1in}}

\newcommand{\f}{\begin{equation}}
\newcommand{\ff}{\end{equation}}
\newcommand{\fa}{\begin{eqnarray}}
\newcommand{\ffa}{\end{eqnarray}}

\begin{document}
\title{The cosmological perturbation theory in loop cosmology with holonomy
corrections}
\author{Jian-Pin Wu}
\email{jianpinwu@yahoo.com.cn} \affiliation{Department of Physics, Beijing Normal University, Beijing 100875, China}
\author{Yi Ling}
\email{yling@ncu.edu.cn} \affiliation{Center for Relativistic Astrophysics and High Energy Physics,Department of Physics, Nanchang University, 330031, China}

\begin{abstract}

In this paper we investigate the scalar mode of first-order metric
perturbations over spatially flat FRW spacetime when the holonomy
correction is taken into account in the semi-classical framework
of loop quantum cosmology. By means of the Hamiltonian derivation,
the cosmological perturbation equations is obtained in
longitudinal gauge. It turns out that in the presence of metric
perturbation the holonomy effects influence both background and
perturbations, and contribute the non-trivial terms $S_{h1}$ and $S_{h2}$ in the
cosmological perturbation equations.

\end{abstract}
\maketitle

\section{Introduction}

In loop quantum cosmology  two main quantum gravity effects lead
to remarkable modifications to the standard description of the
early universe(for a detailed review, see Ref. \cite{ReviewLQC}).
One is due to the holonomy correction and the other is due to the
inverse volume correction. Such modifications can successfully
avoid the Big Bang singularity
\cite{AvoidSingularity,QNBBofATP,QNBBIMofATP,AISS}, and replace it
by the Big Bounce even at the semi-classical level
\cite{BouncingofLQCTM,BouncingofLQCB}. In addition, it is very
interesting to notice that quantum gravity effects may lead to the
occurrence of the super-inflationary phase \cite{Superinflation}.
As shown in Ref. \cite{SuperinflationCFT}, such a
super-inflationary phase can also resolve the horizon problem with
only a few number of e-foldings. Therefore, it is possible to
construct a phase of inflation or an alternative to inflation in
the framework of loop quantum cosmology.

As we all known, the inflationary phase is crucial for
understanding the structure formation and anisotropies of the CMB.
In order to address these issues in the framework of loop quantum
cosmology, we must consider the cosmological perturbation theory
with modifications due to quantum gravity effects. In the earlier
work by Bojowald $et$ $al.$\cite{HamPerturb,FormationBPRL}, by
means of the Hamiltonian derivation they have obtained the
cosmological perturbation equation with inverse volume corrections
for scalar modes in longitudinal gauge. They show that
super-horizon curvature perturbations are not preserved. Recently,
they  have also derived the gauge-invariant quantities and the
corresponding gauge-invariant cosmological perturbation equations
with inverse volume corrections for scalar modes
\cite{AnomalyFreedom,GaugeInvariant}. In addition, the vector
modes and tensor modes with corrections from loop quantum gravity
have been investigated \cite{VectorB,TensorB}.

At the same time, some pioneer work have already been devoted to
understanding the primordial power spectrum in the perturbation
theory of LQC
\cite{SuperinflationCFT,PDPLQC,ISPLQC,ConstraintsLQC,XZandYL,Tsujikawa,Shimano,MJL}. First of
all, in Ref. \cite{SuperinflationCFT,ConstraintsLQC}, it is shown
that a scale invariant spectrum can be obtained. More importantly
these attempts imply that the quantum gravity effects may leave an
imprints on the power spectrum which can be potentially detected
in the future experiments such as the Planck satellite.
However,above considerations are restricted to the scalar field
perturbations with fixed background. To provide a complete and
more precise understanding on the perturbation theory in loop
cosmology, it is essential to take the metric perturbation into
account. Along this direction it is worthwhile to point out that
another potential observables, primordial gravitational waves have
already been investigated intensively in LQC \cite{GWLQC}.

Although, in Ref.
\cite{HamPerturb,FormationBPRL,AnomalyFreedom,GaugeInvariant} the
cosmological perturbation equations with inverse volume
corrections have been derived in longitudinal gauge and
gauge-invariant manner respectively, the metric perturbations with
holonomy corrections is still absent. In the present paper, by
means of the Hamiltonian derivation, we will derive the
cosmological perturbation equations with holonomy corrections in
longitudinal gauge.

The outline of our paper is the following. For comparison, we
firstly present a brief review on the  perturbation equations in
standard classical cosmology in section II.  After introducing the
basic variables in loop cosmology in section III, we will
demonstrate a detailed derivation on the cosmological perturbation
equation with holonomy corrections in section IV. The discussion
is given in section VI.

\section{The classical cosmological perturbation equations}

Before proceeding to the effective loop quantum cosmology with
holonomy corrections, we first briefly review the classical
 perturbation equations in standard cosmology. A detailed derivation can be
found in Ref. \cite{CosmologicalPerturbations}. Let us now
consider a spatially flat background metric of FRW type
\begin{equation} \label{IHFRWM}
ds^{2}=a^{2}(\eta)(-d\eta^{2}+\delta_{ab}dx^{a}dx^{b}) ~.
\end{equation}
where $\eta$ is the conformal time. The spatial part
of the metric describes isotropic and homogeneous $3$-surfaces. Then one can perturb the background metric
\begin{equation} \label{LGFRWM}
ds^{2}=a^{2}(\eta)\left[-(1+2\Phi)d\eta^{2}+(1-2\Psi)\delta_{ab}dx^{a}dx^{b}\right]
~.
\end{equation}
Here we only consider the scalar modes in longitudinal gauge,
which is thus diagonal. Through this paper, we will consider the
scalar field $\varphi$ as the matter source. Expanding the
Einstein's equation linearly, one can obtain the cosmological
perturbation equation
\begin{equation} \label{CPE1}
\nabla^{2}\Phi-3\mathbb{H}\dot{\Phi}-(\dot{\mathbb{H}}+2\mathbb{H}^{2})\Phi=4\pi G(\dot{\bar{\varphi}}\dot{\delta\varphi}+\bar{p}V_{,\bar{\varphi}}(\bar{\varphi})\delta\varphi) ~,
\end{equation}
\begin{equation} \label{CPE2}
\ddot{\Phi}+3\mathbb{H}\dot{\Phi}+(\dot{\mathbb{H}}+2\mathbb{H}^{2})\Phi=4\pi G(\dot{\bar{\varphi}}\dot{\delta\varphi}-\bar{p}V_{,\bar{\varphi}}(\bar{\varphi})\delta\varphi) ~,
\end{equation}
\begin{equation} \label{CPE3}
\partial_{a}(\dot{\Phi}+\mathbb{H}\Phi)=4\pi G\dot{\bar{\varphi}}\delta\varphi_{,a}~,
\end{equation}
where a dot denotes a derivative with respect to the conformal
time $\eta$. $\mathbb{H}$ is the Hubble expansion rate in the
conformal time, and for later convenience, we have identified
$a^{2}$ with $\bar{p}$ which is introduced in (\ref{BV1}). Note
that in the case of vanishing anisotropic stresses, two scalar
functions $\Phi$ and $\Psi$ coincide, $\Phi=\Psi$. Therefore, in
above equations we have set $\Phi=\Psi$, which simplifies the
equations considerably\footnote{However we must point out that, as
a matter of fact, $\Phi=\Psi$ is a consequence of equations of
motion, which can also be seen in this paper}. Moreover, among these equations above only
two of them are independent. Combining
these equations, one can obtain the following second order
differential equation for $\Phi$
\begin{equation} \label{CPE4}
\ddot{\Phi}-\nabla^{2}\Phi+(6\mathbb{H}+2\bar{p}\frac{V_{,\bar{\varphi}}(\bar{\varphi})}{\dot{\bar{\varphi}}})\dot{\Phi}
+(2\dot{\mathbb{H}}+4\mathbb{H}^{2}+\frac{V_{,\bar{\varphi}}(\bar{\varphi})}{\dot{\bar{\varphi}}}\mathbb{H})\Phi=0~.
\end{equation}

In addition, the background and the perturbed Klein-Gordon
equation can respectively expressed as
\begin{equation} \label{BKG}
\ddot{\bar{\varphi}}+2\mathbb{H}\dot{\bar{\varphi}}+\bar{p}V_{,\varphi}(\bar{\varphi})=0~,
\end{equation}
\begin{equation} \label{PKG}
\ddot{\delta\varphi}+2\mathbb{H}\dot{\delta\varphi}-\bigtriangledown^{2}\delta\varphi+\bar{p}V_{,\bar{\varphi}\bar{\varphi}}(\bar{\varphi})
+2\bar{p}V_{,\bar{\varphi}}(\bar{\varphi})\Phi-4\dot{\bar{\varphi}}\dot{\Phi}=0~.
\end{equation}

\section{The basic variables}

Now we intend to study the scalar mode of first-order metric
perturbations around spatially flat FRW spacetime when the
holonomy corrections is taken into account in the semi-classical
framework of loop quantum cosmology. To derive the cosmological
perturbation equations we adopt the Hamiltonian approach which has
been developed in the effective loop quantum cosmology with
inverse triad corrections \cite{HamPerturb,GaugeInvariant}. We
summarize the basic idea and steps as follows.

In loop quantum gravity, instead of the spatial metric $q_{ab}$, a
densitized triad $E^{a}_{i}$ is primarily used, which satisfies
$E^{a}_{i}E^{b}_{i}=q^{ab}detq$. Moreover, in the canonical
formulation the space-time metric is given by
\begin{equation} \label{CSTM}
ds^{2}=-N^{2}d\eta^{2}+q_{ab}(dx^{a}+N^{a}d\eta)(dx^{b}+N^{b}d\eta) ~,
\end{equation}
where $N$ and $N^{a}$ are lapse function and shift vector
respectively.

By comparing the above equation with the FRW metric
(\ref{IHFRWM}), the background variables, $\bar{N}$, $\bar{N^{a}}$ and
$\bar{E}^{a}_{i}$, can be expressed as respectively
\begin{equation} \label{BV1}
\bar{N}=\sqrt{\bar{p}}; \bar{N^{a}}=0; \bar{E}^{a}_{i}=\bar{p}\delta^{a}_{i} ~,
\end{equation}
where the background variables are denoted with a bar, which
describe smoothed out, spatial averaged quantities. Another
background variable, the extrinsic curvature components
$\bar{K}^{i}_{a}$, can be derived from the following relation
\begin{equation} \label{EC1}
\bar{K}_{ab}=\frac{1}{2\bar{N}}(\dot{\bar{q}}_{ab}-2D_{(a}\bar{N}_{b)})=\dot{a}\delta_{ab} ~.
\end{equation}
where D is the covariant spatial derivation. Thus, the extrinsic curvature can be expressed as
\begin{equation} \label{BV2}
\bar{K}^{i}_{a}=\frac{\bar{E}^{b}_{i}}{\sqrt{|det(\bar{E}^{c}_{j})|}}\bar{K}_{ab}=\frac{\dot{\bar{p}}}{2\bar{p}}\delta^{i}_{a}=:\bar{k}\delta^{i}_{a} ~.
\end{equation}

In equation (\ref{BV2}), we have defined the background extrinsic
curvature as
$\bar{k}=:\frac{\dot{\bar{p}}}{2\bar{p}}=\frac{\dot{a}}{a}$, which can
also be obtained from the background equations of motion
\cite{GaugeInvariant}. Therefore, in classical FRW background, the
extrinsic curvature is nothing but the conformal Hubble parameter
$\mathbb{H}$. However, in the effective loop quantum cosmology,
the relation between the extrinsic curvature and the conformal
Hubble parameter will change due to quantum gravity corrections,
which we will see in the next section.

The canonical perturbed variables can be related to the perturbed
metric variables by comparing the perturbed metric (\ref{LGFRWM})
with the canonical one(\ref{CSTM}). It turns out that the
perturbed triad is given by
\begin{equation} \label{BPtriad2}
\delta E^{a}_{i}=-2\bar{p}\Psi\delta^{a}_{i}~,
\end{equation}

and the perturbed lapse function is
\begin{equation} \label{BPls}
\delta N=\bar{N}\Phi.
\end{equation}

As shown in the above, the extrinsic curvature components can be diagonal, thus it
can be expanded as
\begin{equation} \label{BPcurvature1}
K^{i}_{a}=\bar{K}^{i}_{a}+\delta K^{i}_{a}=\bar{k}\delta^{i}_{a}+\delta K^{i}_{a} ~.
\end{equation}

The perturbed extrinsic curvature will be derived from the
equation of motion in the following. We can assume that $\delta E^{a}_{i}$ and $\delta K^{i}_{a}$ do not have
homogeneous modes, namely
\begin{equation} \label{PHmode}
\int_{\Sigma} \delta E^{a}_{i}\delta^{i}_{a}d^{3}x=0, \int_{\Sigma} \delta K^{i}_{a}\delta^{a}_{i}d^{3}x=0~.
\end{equation}

And the homogeneous mode is defined by
\begin{equation} \label{Hmode}
\bar{p}=\frac{1}{3V_{0}}\int_{\Sigma} E^{a}_{i}\delta^{i}_{a}d^{3}x, \bar{k}=\frac{1}{3V_{0}}\int_{\Sigma} K^{i}_{a}\delta^{a}_{i}d^{3}x~,
\end{equation}
where we integrate over a bounded region of coordinate size
$V_{0}=\int_{\Sigma} d^{3}x$. Then we can construct the Poisson
brackets of the background and perturbed variables
\cite{AnomalyFreedom},
\begin{equation} \label{PB}
\{\bar{k},\bar{p}\}=\frac{8\pi G}{3V_{0}}, \{\delta K^{i}_{a}(x),\delta E^{b}_{j}(y)\}=8\pi G\delta^{i}_{j}\delta^{b}_{a}\delta^{3}(x-y)~.
\end{equation}

In addition, we point out that the similar conditions will be
required in the perturbed lapse $\delta N$, the scalar field
$\delta \varphi$ and conjugate momentum $\delta \pi$ such that
\begin{equation} \label{PHmode1}
\int_{\Sigma} \delta N d^{3}x=0, \int_{\Sigma} \delta \varphi d^{3}x=0, \int_{\Sigma} \delta \pi d^{3}x=0~,
\end{equation}
which is used in expanding the Hamiltonian constraint. While the
homogeneous mode of the scalar field and its conjugate momentum is
\begin{equation} \label{Hmode1}
\bar{\varphi}=\frac{1}{V_{0}}\int_{\Sigma} \varphi d^{3}x, \bar{\pi}=\frac{1}{V_{0}}\int_{\Sigma} \pi d^{3}x~.
\end{equation}

Therefore, the Poisson brackets of the background and perturbed variables of scalar field is
\begin{equation} \label{PBS}
\{\bar{\varphi},\bar{\pi}\}=\frac{1}{3V_{0}}, \{\delta \varphi(x),\delta \pi(y)\}=\delta^{3}(x-y)~.
\end{equation}

\section{The cosmological perturbation theory with holonomy corrections}

Now we turn to the derivation of the cosmological perturbation
theory in the effective loop quantum cosmology with holonomy
corrections. For more details on the Hamilton cosmological
perturbation theory, we refer to
Ref.\cite{HamPerturb,GaugeInvariant}.

Thanks to the holonomy corrections, in the isotropic and
homogeneous models, the effective Hamiltonian can be
obtained at the phenomenological level by simply replacing the
background Ashtekar connection $\bar{k}$ by $\frac{\sin
\bar{\mu}\gamma \bar{k}}{\bar{\mu}\gamma}$, where $\gamma$ is the
Barbero-Immirzi parameter. The parameter $\bar{\mu}$ depends on
the quantization scheme and may be a function of $\bar{p}$. More discussions about
the parameter $\bar{\mu}$, we can refer to Ref. \cite{QNBBIMofATP,muScheme}.

However, when the inhomogeneities are taken into account, it is no longer true. To
study the effects of holonomy corrections on inhomogeneous
perturbations, we similarly substitute the appearance of $\bar{k}$ in
the classical Hamiltonian by a general form $\frac{\sin m
\bar{\mu}\gamma \bar{k}}{m\bar{\mu}\gamma}$ where $m$ is an
integer. In the context of vector modes \cite{VectorB} and tensor
modes \cite{TensorB}, due to the requirement of the anomaly
cancellation, we can fix the parameter $m$. Since the evolution of all modes
should be generated by one general Hamiltonian constraint, it would be
reasonable to use the values found for vector modes and tensor modes also for
scalars. However, it must been also pointed out that the restrictions of anomaly
cancellation from the vector modes and tensor
modes has not been checked for scalars in the presence of
holonomy corrections. Complete consistency is realized only if all modes can be
anomaly-free with holonomy corrections for specific parameter values.

Subsequently, one can write down the expressions for the
gravitational Hamiltonian density in a similar manner ${\cal H}_{G}^{h}={\cal
H}_{G}^{h(0)}+ {\cal H}_{G}^{h(1)}+{\cal H}_{G}^{h(2)}$ with
\begin{eqnarray} \label{HC0}
{\mathcal H}_{G}^{h(0)} &=& -6(\frac{\sin \bar{\mu}\gamma \bar{k}}{\bar{\mu}\gamma})^{2}\sqrt{\bar p}~,
\nonumber\\ \label{HC1} {\mathcal H}_{G}^{h(1)} &=&
-4 (\frac{\sin 2\bar{\mu}\gamma \bar{k}}{2\bar{\mu}\gamma})\sqrt{\bar{p}} \delta^c_j\delta K_c^j
-\frac{1}{\sqrt{\bar{p}}}(\frac{\sin \bar{\mu}\gamma \bar{k}}{\bar{\mu}\gamma})^{2} \delta_c^j\delta E^c_j
+\frac{2}{\sqrt{\bar{p}}}
\partial_c\partial^j\delta E^c_j  ~,
\nonumber\\ \label{HC2} {\mathcal H}_{G}^{h(2)} &=&
\sqrt{\bar{p}} \delta K_c^j\delta K_d^k\delta^c_k\delta^d_j -
\sqrt{\bar{p}} (\delta K_c^j\delta^c_j)^2
-\frac{2}{\sqrt{\bar{p}}}(\frac{\sin 2\bar{\mu}\gamma \bar{k}}{2\bar{\mu}\gamma}) \delta E^c_j\delta K_c^j-\frac{1}{2\bar{p}^{3/2}}(\frac{\sin \bar{\mu}\gamma \bar{k}}{\bar{\mu}\gamma})^{2} \delta E^c_j\delta
E^d_k\delta_c^k\delta_d^j
\nonumber\\
&& \quad
+\frac{1}{4\bar{p}^{3/2}}(\frac{\sin \bar{\mu}\gamma \bar{k}}{\bar{\mu}\gamma})^{2}(\delta E^c_j\delta_c^j)^2
-\frac{\delta^{jk} }{2\bar{p}^{3/2}}(\partial_c\delta E^c_j)
(\partial_d\delta E^d_k)  ~,
\end{eqnarray}
where the superscript $``h"$ represents the holonomy corrections
and the corresponding classical expressions can be found in
Ref.\cite{AnomalyFreedom,GaugeInvariant}. We now only consider the
scalar field as the matter source. Its Hamiltonian density expands
as ${\cal H}_{M}={\cal H}_{M}^{(0)}+ {\cal H}_{M}^{(1)}+{\cal
H}_{M}^{(2)}$. Since the matter is free from the holonomy
corrections, the expressions of scalar field Hamiltonian density,
${\cal H}_{M}={\cal H}_{\pi}+ {\cal H}_{\nabla}+{\cal
H}_{\varphi}$, expanding up to the second order, are as the
classical cases \cite{AnomalyFreedom,GaugeInvariant},
\begin{equation}
\label{SFPE0}
\mathcal{H}_{\pi}^{(0)}=\frac{\bar{\pi}^{2}}{2\bar{p}^{3/2}}, \mathcal{H}_{\nabla}^{(0)}=0, \mathcal{H}_{\varphi}^{(0)}=\bar{p}^{3/2}V(\bar{\varphi}),
\end{equation}
\begin{equation}
\label{SFPE1}
\mathcal{H}_{\pi}^{(1)}=\frac{\bar{\pi}\delta\pi}{\bar{p}^{3/2}}-\frac{\bar{\pi}^{2}}{2\bar{p}^{3/2}}\frac{\delta^{j}_{c}\delta E^{c}_{j}}{2\bar{p}}, \mathcal{H}_{\nabla}^{(1)}=0, \mathcal{H}_{\varphi}^{(1)}=\bar{p}^{3/2}(V_{,\bar{\varphi}}(\bar{\varphi})\delta \varphi+V(\bar{\varphi})\frac{\delta^{j}_{c}\delta E^{c}_{j}}{2\bar{p}}),
\end{equation}
and
\begin{eqnarray} \label{SFPE2}
\mathcal{H}_{\pi}^{(2)} &=& \frac{1}{2}\frac{\delta\pi^{2}}{\bar{p}^{3/2}}
-\frac{\bar{\pi}\delta \pi}{\bar{p}^{3/2}}\frac{\delta^{j}_{c}\delta E^{c}_{j}}{2\bar{p}}
+\frac{1}{2}\frac{\bar{\pi}^{2}}{\bar{p}^{3/2}}\left(\frac{(\delta^{j}_{c}\delta E^{c}_{j})^{2}}{8\bar{p}^{2}}+\frac{\delta^{k}_{c}\delta^{j}_{d}\delta E^{c}_{j}\delta E^{d}_{k}}{4\bar{p}^{2}}\right)~,
\nonumber\\ \label{SFPE21} \mathcal{H}_{\nabla}^{(2)} &=& \frac{1}{2}\sqrt{\bar{p}}\delta^{ab}\partial_{a}\delta \varphi \partial_{b}\delta \varphi~,
\nonumber\\ \label{SFPE22} \mathcal{H}_{\varphi}^{(2)} &=& \frac{1}{2}\bar{p}^{3/2}V_{,\bar{\varphi}\bar{\varphi}}(\bar{\varphi})\delta \varphi^{2}+\bar{p}^{3/2}V_{,\bar{\varphi}}(\bar{\varphi})\delta \varphi\frac{\delta^{j}_{c}\delta E^{c}_{j}}{2\bar{p}}
\nonumber\\
&& \quad
+\bar{p}^{3/2}V(\bar{\varphi})\left(\frac{(\delta^{j}_{c}\delta E^{c}_{j})^{2}}{8\bar{p}^{2}}-\frac{\delta^{k}_{c}\delta^{j}_{d}\delta E^{c}_{j}\delta E^{d}_{k}}{4\bar{p}^{2}}\right)  ~.
\end{eqnarray}

\subsection{The background equations}

In the isotropic and homogeneous FRW background, the diffeomorphism
constraint vanishes. Therefore background equations are generated
only by the background Hamiltonian constraint, which can be
expressed as
\begin{equation} \label{BHC0}
H^{h(0)}[\bar{N}]=\frac{1}{16\pi G}\int_{\Sigma}d^{3}x\bar{N}[\mathcal{H}_{G}^{(0)}+16\pi G(\mathcal{H}_{\pi}^{(0)}+\mathcal{H}_{\varphi}^{(0)})] ~.
\end{equation}

Thus the explicit expression for the background Hamiltonian constraint is

%
\begin{equation} \label{BCE}
-\frac{3}{8\pi G}\sqrt{\bar{p}}(\frac{\sin \bar{\mu}\gamma \bar{k}}{\bar{\mu}\gamma})^{2}+\frac{\bar{\pi}^{2}}{2\bar{p}^{3/2}}+\bar{p}^{3/2}V(\bar{\varphi})=0~.
\end{equation}

Then, by means of Poisson bracket, we can derive the equation of
motion for the gravitational variables $\bar{k}$ and $\bar{p}$.
\begin{equation} \label{HEOMk}
\dot{\bar{k}}=\{\bar{k},H^{h(0)}[\bar{N}]\}=-[\frac{1}{2}(\frac{\sin \bar{\mu}\gamma \bar{k}}{\bar{\mu}\gamma})^{2}+\bar{p}\frac{\partial}{\partial \bar{p}}(\frac{\sin \bar{\mu}\gamma \bar{k}}{\bar{\mu}\gamma})^{2}]+4\pi G[-\frac{\bar{\pi}^{2}}{2\bar{p}^{2}}+\bar{p}V(\bar{\varphi})] ~.
\end{equation}
\begin{equation} \label{HEOMp}
\dot{\bar{p}}=\{\bar{p},H^{h(0)}[\bar{N}]\}=2\bar{p}(\frac{\sin 2\bar{\mu}\gamma \bar{k}}{2\bar{\mu}\gamma}) ~.
\end{equation}

Similarly, the equation of motion for scalar field $\bar{\varphi}$
and its conjugate momentum field $\bar{\pi}$ can also be derived as
\begin{equation} \label{HEOMvp}
\dot{\bar{\varphi}}=\{\bar{\varphi},H^{h(0)}[\bar{N}]\}=\frac{\bar{\pi}}{\bar{p}} ~.
\end{equation}
\begin{equation} \label{HEOMpi}
\dot{\bar{\pi}}=\{\bar{\pi},H^{h(0)}[\bar{N}]\}=-\bar{p}^{2}V_{,\bar{\varphi}}(\bar{\varphi}) ~.
\end{equation}

Note that in above Poisson brackets, we have used the relation
$\bar{N}=\sqrt{\bar{p}}$. Substituting the relation (\ref{HEOMvp})
into the constraint equation (\ref{BCE}) gives rise to the corrected
Friedmann equation
\begin{equation} \label{CFRWE}
(\frac{\sin \bar{\mu}\gamma \bar{k}}{\bar{\mu}\gamma})^{2}=\frac{8\pi G}{3}[\frac{1}{2}\dot{\bar{\varphi}}^{2}+\bar{p}V(\varphi)]~.
\end{equation}

At the same time, equation (\ref{HEOMk}) is just the corrected
Raychaudhuri equation
\begin{equation} \label{CRayE}
\dot{\bar{k}}+\frac{1}{2}(\frac{\sin \bar{\mu}\gamma \bar{k}}{\bar{\mu}\gamma})^{2}+\bar{p}\frac{\partial}{\partial \bar{p}}(\frac{\sin \bar{\mu}\gamma \bar{k}}{\bar{\mu}\gamma})^{2}=4\pi G[-\frac{\dot{\bar{\varphi}}^{2}}{2}+\bar{p}V(\bar{\varphi})] ~.
\end{equation}

In the classical limit, $\bar{\mu}\rightarrow 0$, above two
equations can be reduced to the Friedmann and Raychaudhuri
equation in the standard cosmology. Finally, the Klein-Gordon
equation can be derived from Eqs. (\ref{HEOMvp}), (\ref{HEOMpi})
and (\ref{HEOMp})
\begin{equation} \label{KG}
\ddot{\bar{\varphi}}+2(\frac{\sin 2\bar{\mu}\gamma \bar{k}}{2\bar{\mu}\gamma})\dot{\bar{\varphi}}+\bar{p}V_{,\bar{\varphi}}(\bar{\varphi})=0~.
\end{equation}

In addition, from the equation of motion (\ref{HEOMp}), one can find
that the extrinsic curvature $\bar{k}$ is related to the conformal
Hubble parameter $\mathbb{H}$ by
\begin{equation} \label{RECandHP}
\frac{\sin 2\bar{\mu}\gamma \bar{k}}{2\bar{\mu}\gamma}=\frac{\dot{\bar{p}}}{2\bar{p}}=:\mathbb{H}~.
\end{equation}

Therefore, due to the holonomy corrections, the conformal Hubble
parameter $\mathbb{H}$ is not simply equal to the extrinsic
curvature $\bar{k}$ but receives corrections. 
For consistency, in our next derivation we will continuously use the
extrinsic curvature $\bar{k}$ rather than the conformal Hubble
parameter. Only at the end, we will use the conformal Hubble
parameter $\mathbb{H}$ instead of $\frac{\sin 2\bar{\mu}\gamma
\bar{k}}{2\bar{\mu}\gamma}$ in the perturbation equations.

\subsection{The perturbed equations}

In this subsection, we will derive the cosmological perturbation
equation with holonomy corrections. Firstly we will derive the
equations of motion of perturbed variables. In the canonical
formulation, the equation of motion of any phase space function $f$
is determined by Poisson bracket, $\dot{f}=\{f,H\}$. Here $H$ is the
total Hamiltonian, which is a sum of the Hamiltonian constraint
$H[N]$ and the diffeomorphism constraint $D[N^{a}]$,
$H=H[N]+D[N^{a}]$. Since the zero-order and first-order shift
vectors vanish,  the diffeomorphism constraints is identically
satisfied up to the second-order. Thus, the equations of motion of
the perturbed variables are only generated by the Hamiltonian
constraint. The perturbed Hamiltonian constraint up to the
second-order is written as
$\tilde{H}^{h}[N]=\tilde{H}^{h}[\bar{N}]+\tilde{H}^{h}[\delta N]$
with
\begin{eqnarray} \label{BHC0}
\tilde{H}^{h}[\bar{N}]&=&\frac{1}{16\pi G}\int_{\Sigma}d^{3}x\bar{N}[\mathcal{H}_{G}^{(2)}+16\pi G(\mathcal{H}_{\pi}^{(2)}+\mathcal{H}_{\nabla}^{(2)}+\mathcal{H}_{\varphi}^{(2)})]~,
\nonumber\\ \label{BHC01} \tilde{H}^{h}[\delta N] &=&\frac{1}{16\pi G}\int_{\Sigma}d^{3}x\delta N[\mathcal{H}_{G}^{(1)}+16\pi G(\mathcal{H}_{\pi}^{(1)}+\mathcal{H}_{\varphi}^{(1)})]~.
\end{eqnarray}

Note that we have used the conditions that the perturbed variables
do not have homogeneous modes as described in Eq.(\ref{PHmode})
and (\ref{PHmode1}). As well, we input the boundary condition
requiring that the integration over the boundary vanishes, namely
\begin{equation} \label{INT0}
\int_{\Sigma}\bar{N}[\mathcal{H}_{G}^{h1}+16\pi G (\mathcal{H}_{\pi}^{1}+\mathcal{H}_{\varphi}^{1})] =0, \int_{\Sigma}\delta N[\mathcal{H}_{G}^{h0}+16\pi G (\mathcal{H}_{\pi}^{0}+\mathcal{H}_{\varphi}^{0})]=0~.
\end{equation}

Therefore, the equations of motion of perturbed variables are
generated only by the second order part of Hamiltonian constraints.
Thus, we can arrive at the equation of motion of the perturbed
variables by means of the Poisson bracket
\begin{eqnarray} \label{PEDK}
\delta \dot{K}^{i}_{a} &\equiv& \{\delta K^{i}_{a},\tilde{H}^{h}[\bar{N}]+\tilde{H}^{h}[\delta N]\}
\nonumber\\ \label{PEDK1} &=& \frac{\bar{N}}{\bar{p}^{3/2}}[-\bar{p}(\frac{\sin 2\bar{\mu}\gamma \bar{k}}{2\bar{\mu}\gamma})\delta K^{i}_{a}-\frac{1}{2}(\frac{\sin \bar{\mu}\gamma \bar{k}}{\bar{\mu}\gamma})^{2}\delta E^{d}_{k}\delta^{k}_{a}\delta^{i}_{d}+\frac{1}{4}(\frac{\sin \bar{\mu}\gamma \bar{k}}{\bar{\mu}\gamma})^{2}(\delta E^{d}_{k}\delta^{k}_{d})\delta^{i}_{a}+\frac{1}{2}\delta^{ik}\partial_{a}\partial_{d}\delta E^{d}_{k}]
\nonumber\\
&& \quad
-\frac{1}{2}\frac{\delta N}{\sqrt{\bar{p}}}(\frac{\sin \bar{\mu}\gamma \bar{k}}{\bar{\mu}\gamma})^{2}\delta^{i}_{a}+\frac{1}{\sqrt{\bar{p}}}\partial_{a}\partial^{i}\delta N
\nonumber\\
&& \quad
+4\pi G\frac{\bar{N}}{\bar{p}^{3/2}}[-\frac{\bar{\pi}\delta \pi}{\bar{p}}\delta^{i}_{a}+\frac{1}{2}\frac{\bar{\pi}^{2}}{\bar{p}^{2}}(\frac{1}{2}\delta E^{d}_{k}\delta^{k}_{d}\delta^{i}_{a}+\delta E^{d}_{k}\delta^{k}_{a}\delta^{i}_{d})+\bar{p}^{2}V_{,\bar{\varphi}}(\bar{\varphi})\delta \varphi \delta^{i}_{a}
\nonumber\\
&& \quad
+\bar{p}V(\varphi)(\frac{1}{2}\delta E^{d}_{k}\delta^{k}_{d}\delta^{i}_{a}-\delta E^{d}_{k}\delta^{k}_{a}\delta^{i}_{d})]+4\pi G\delta N[-\frac{1}{2}\frac{\bar{\pi}^{2}}{\bar{p}^{5/2}}+\sqrt{\bar{p}}V(\bar{\varphi})]\delta^{i}_{a},
\end{eqnarray}
\begin{eqnarray} \label{PEDE}
\dot{\delta E^{a}_{i}} &\equiv& \{\delta E^{a}_{i},\tilde{H}^{h}[\bar{N}]+\tilde{H}^{h}[\delta N]\}
\nonumber\\ \label{PEDE1} &=&\frac{\bar{N}}{\sqrt{\bar{p}}}[-\bar{p}\delta K^{j}_{c}\delta^{c}_{i}\delta^{a}_{j}+\bar{p}(\delta K^{j}_{c}\delta^{c}_{j})\delta^{a}_{i}+(\frac{\sin 2\bar{\mu}\gamma \bar{k}}{2\bar{\mu}\gamma})\delta E^{a}_{i}]+2\delta N (\frac{\sin 2\bar{\mu}\gamma \bar{k}}{2\bar{\mu}\gamma})\sqrt{\bar{p}}\delta^{a}_{i},
\end{eqnarray}
\begin{equation} \label{PEDS}
\delta\dot{\varphi}\equiv \{\delta\varphi,\tilde{H}^{h}[\bar{N}]+\tilde{H}^{h}[\delta N]\}=\frac{\bar{N}}{\bar{p}^{3/2}}(\delta \pi -\bar{\pi}\frac{\delta E^{c}_{j}\delta^{j}_{c}}{2\bar{p}})+\frac{\delta N}{\bar{p}^{3/2}}\bar{\pi}~,
\end{equation}
\begin{equation} \label{PEDpi}
\delta\dot{\pi}\equiv \{\delta\pi,\tilde{H}^{h}[\bar{N}]+\tilde{H}^{h}[\delta N]\}=\frac{\bar{N}}{\bar{p}^{3/2}}
[\bar{p}^{2}\nabla^{2}\delta\varphi-\bar{p}^{3}V_{,\bar{\varphi}\bar{\varphi}}\delta\varphi
-\frac{1}{2}\bar{p}^{2}V_{,\bar{\varphi}\bar{\varphi}}\delta E^{c}_{j}\delta^{j}_{c}]~.
\end{equation}

Furthermore, using Eqs.(\ref{BPtriad2}) and (\ref{HEOMp}),  we can
obtain the perturbed extrinsic curvature $\delta K^{i}_{a}$ from
equation (\ref{PEDE}),
\begin{equation} \label{BPcurvature2}
\delta K^{i}_{a}=-\delta^{i}_{a}[\dot{\Psi}+(\frac{\sin 2\bar{\mu}\gamma \bar{k}}{2\bar{\mu}\gamma})(\Psi+\Phi)]~.
\end{equation}

Similarly, using Eq.(\ref{BPtriad2}), equations (\ref{PEDS}) and
(\ref{PEDpi}) can be respectively reexpressed as
\begin{equation} \label{PEDS1}
\delta\dot{\varphi}=\frac{\delta \pi}{\bar{p}}+\frac{\bar{\pi}}{\bar{p}}(3\Psi+\Phi)~.
\end{equation}
\begin{equation} \label{PEDpi1}
\delta\dot{\pi}=\bar{p}\nabla^{2}\delta\varphi-\bar{p}^{2}V_{,\bar{\varphi}\bar{\varphi}}\delta\varphi
+3\bar{p}^{2}V_{,\bar{\varphi}}\Psi~.
\end{equation}

Now, we derive the Hamiltonian's equation using the equation of
motion of $\delta K^{i}_{a}$. Collecting the expressions $\delta
E^{a}_{i}$ (\ref{BPtriad2}), $\delta K^{i}_{a}$ (\ref{BPcurvature2})
, $\delta N$(\ref{BPls}), and equations (\ref{HEOMvp}),
(\ref{PEDS1}), one can obtain
\begin{eqnarray} \label{HCE0}
\{\ddot{\Psi}&+&(\frac{\sin 2\bar{\mu}\gamma \bar{k}}{2\bar{\mu}\gamma})(2\dot{\Psi}+\dot{\Phi})
+[(\cos2\bar{\mu}\gamma \bar{k}-\frac{1}{2})\dot{\bar{k}}+(\frac{\sin 2\bar{\mu}\gamma \bar{k}}{2\bar{\mu}\gamma})^{2}+\frac{\dot{\bar{\mu}}}{\bar{\mu}}(\bar{k}\cos2\bar{\mu}\gamma \bar{k}-\frac{\sin 2\bar{\mu}\gamma \bar{k}}{2\bar{\mu}\gamma})
\nonumber\\
&& \quad
-\frac{1}{2}\bar{p}\frac{\partial}{\partial \bar{p}}(\frac{\sin \bar{\mu}\gamma \bar{k}}{\bar{\mu}\gamma})^{2}](\Psi+\Phi)+\bar{p}V(\bar{\varphi})(\Phi-\Psi)\}\delta^{i}_{a}
+\partial_{a}\partial^{i}(\Phi-\Psi)
\nonumber\\
&& \quad
=4\pi G(\dot{\bar{\varphi}}\dot{\delta \varphi}-\bar{p}V_{,\bar{\varphi}}(\bar{\varphi})\delta \varphi) ~.
\end{eqnarray}

When deriving this equation, we have used the relation
\begin{equation} \label{HCE}
4\pi G\dot{\bar{\varphi}}^{2}=(\frac{\sin \bar{\mu}\gamma \bar{k}}{\bar{\mu}\gamma})^{2}-\dot{\bar{k}}-\bar{p}\frac{\partial}{\partial \bar{p}}(\frac{\sin \bar{\mu}\gamma \bar{k}}{\bar{\mu}\gamma})^{2}~,
\end{equation}
which can be obtained from the corrected Friedmann equation
(\ref{CFRWE}) and Raychaudhuri equation (\ref{CRayE}). From equation
(\ref{HCE0}), we can read the off-diagonal equation
\begin{equation} \label{ODE}
\partial_{a}\partial^{i}[\Phi-\Psi]=0~,
\end{equation}
which implies $\Phi=\Psi$. Therefore, in the following derivation, we will
identify $\Phi$ with $\Psi$. Then the diagonal equation gives
\begin{eqnarray} \label{HCE1}
\ddot{\Phi}&+&3(\frac{\sin 2\bar{\mu}\gamma \bar{k}}{2\bar{\mu}\gamma})\dot{\Phi}
+[(2\cos2\bar{\mu}\gamma \bar{k}-1)\dot{\bar{k}}+2(\frac{\sin 2\bar{\mu}\gamma \bar{k}}{2\bar{\mu}\gamma})^{2}+2\frac{\dot{\bar{\mu}}}{\bar{\mu}}(\bar{k}\cos2\bar{\mu}\gamma \bar{k}-\frac{\sin 2\bar{\mu}\gamma \bar{k}}{2\bar{\mu}\gamma})
\nonumber\\
&& \quad
-\bar{p}\frac{\partial}{\partial \bar{p}}(\frac{\sin \bar{\mu}\gamma \bar{k}}{\bar{\mu}\gamma})^{2}]\Phi
=4\pi G(\dot{\bar{\varphi}}\dot{\delta \varphi}-\bar{p}V_{,\bar{\varphi}}(\bar{\varphi})\delta \varphi) ~.
\end{eqnarray}

Subsequently, we will consider the diffeomorphism constraint
equation. The perturbed diffeomophism constraint with
holonomy corrections is
\begin{equation} \label{DCE}
D[N^{c}]=\frac{1}{8\pi G}\int_{\Sigma}d^{3}x\delta N^{c}[\bar{p}\partial_{c}(\delta^{d}_{k}\delta K^{k}_{d})-\bar{p}(\partial_{k}\delta K^{k}_{c})-(\frac{\sin 2\bar{\mu}\gamma \bar{k}}{2\bar{\mu}\gamma})\delta^{k}_{c}(\partial_{d}\delta E^{d}_{k})+8\pi G \bar{\pi}\partial_{c}\delta \varphi]~.
\end{equation}

The diffeomorphism constraint equation can be obtained by varying
the diffeomorphism constraint with respect to the shift
perturbation:
\begin{equation} \label{DCE0}
8\pi G\frac{\delta D[\delta N^{c}]}{\delta (\delta N^{c})}=\bar{p}\partial_{c}(\delta^{d}_{k}\delta K^{k}_{d})-\bar{p}(\partial_{k}\delta K^{k}_{c})-(\frac{\sin 2\bar{\mu}\gamma \bar{k}}{2\bar{\mu}\gamma})\delta^{k}_{c}(\partial_{d}\delta E^{d}_{k})+8\pi G \bar{\pi}\partial_{c}\delta \varphi=0~.
\end{equation}

Using the expressions $\delta E^{a}_{i}$ (\ref{BPtriad2}), $\delta
K^{i}_{a}$ (\ref{BPcurvature2}) and equation (\ref{HEOMvp}), the
above equation reduces to
\begin{equation} \label{DCE1}
\partial_{c}[\dot{\Phi}+(\frac{\sin 2\bar{\mu}\gamma \bar{k}}{2\bar{\mu}\gamma})\Phi]=4\pi G \dot{\bar{\varphi}}\partial_{c}\delta\varphi~.
\end{equation}

Finally, we will derive the Hamiltonian constraint equation. We note
that after the variation with respect to the background lapse
$\bar{N}$, the constraint equation will be second-order and can be
neglected. So one can obtain the Hamiltonian constraint equation by
only varying the perturbed lapse $\delta N$
\begin{eqnarray} \label{HCS}
\frac{\delta \tilde{H}^{h}[N]}{\delta(\delta N)}&=&\frac{1}{16\pi G}[-4\frac{\sin 2\bar{\mu} \gamma \bar{k}}{2\bar{\mu} \gamma \bar{k}}\sqrt{\bar{p}}\delta K^{i}_{a}\delta^{a}_{i}-(\frac{\sin \bar{\mu} \gamma \bar{k}}{\bar{\mu} \gamma \bar{k}})^{2}\frac{1}{\sqrt{\bar{p}}}\delta E^{a}_{i}\delta^{i}_{a}+\frac{2}{\sqrt{\bar{p}}}\partial_{a}\partial^{i}\delta E^{a}_{i}]
\nonumber\\
&& \quad
+\frac{\bar{\pi}\delta\pi}{\bar{p}^{3/2}}-(\frac{\bar{\pi}^{2}}{2\bar{p}^{3/2}}
-\bar{p}^{3/2}V(\bar{\varphi}))\frac{\delta E^{a}_{i}\delta^{i}_{a}}{2\bar{p}}+\bar{p}^{3/2}V_{,\bar{\varphi}}(\bar{\varphi})\delta\varphi
\nonumber\\
&& \quad
=0~.
\end{eqnarray}

Substituting the expressions $\delta E^{a}_{i}$ (\ref{BPtriad2}),
$\delta K^{i}_{a}$ (\ref{BPcurvature2}) and equation (\ref{HEOMvp})
into the above equation yields the Hamilton constraint equation
\begin{equation} \label{HCE}
\nabla^{2}\Phi -3(\frac{\sin 2\bar{\mu}\gamma \bar{k}}{2\bar{\mu}\gamma})\dot{\Phi}-[\dot{\bar{k}}+6(\frac{\sin 2\bar{\mu}\gamma \bar{k}}{2\bar{\mu}\gamma})^{2}-4(\frac{\sin \bar{\mu}\gamma \bar{k}}{\bar{\mu}\gamma})^{2}+\bar{p}\frac{\partial}{\partial \bar{p}}(\frac{\sin \bar{\mu}\gamma \bar{k}}{\bar{\mu}\gamma})^{2}]\Phi=4\pi G[\dot{\bar{\varphi}}\dot{\delta \varphi}+\bar{p}V_{,\bar{\varphi}}(\bar{\varphi})\delta \varphi]~.
\end{equation}

In addition, using Eqs.(\ref{PEDS1}) and (\ref{PEDpi1}), with the
help of the background equations (\ref{HEOMp}), (\ref{HEOMvp}) and
(\ref{HEOMpi}), the perturbed Klein-Gordon equation can be expressed
as
\begin{equation} \label{PKG}
\delta\ddot{\varphi}+2(\frac{\sin 2\bar{\mu}\gamma \bar{k}}{2\bar{\mu}\gamma})\delta\dot{\varphi}-\nabla^{2}\delta\varphi+\bar{p}V_{,\bar{\varphi}\bar{\varphi}}\delta\varphi
+2\bar{p}V_{,\bar{\varphi}}\Phi-4\dot{\bar{\varphi}}\dot{\Phi}=0~.
\end{equation}

Now, we replace $\frac{\sin 2\bar{\mu}\gamma
\bar{k}}{2\bar{\mu}\gamma}$ by Hubble parameter $\mathbb{H}$ in the
perturbation equations (\ref{HCE}), (\ref{HCE1}) and (\ref{DCE1})
such that these equations can be reexpressed as
\begin{equation} \label{LCPE1}
\nabla^{2}\Phi-3\mathbb{H}\dot{\Phi}-[\dot{\bar{k}}+6\mathbb{H}^{2}-4(\frac{\sin \bar{\mu}\gamma \bar{k}}{\bar{\mu}\gamma})^{2}+\bar{p}\frac{\partial}{\partial \bar{p}}(\frac{\sin \bar{\mu}\gamma \bar{k}}{\bar{\mu}\gamma})^{2}]\Phi=4\pi G(\dot{\bar{\varphi}}\dot{\delta\varphi}+\bar{p}V_{,\bar{\varphi}}(\bar{\varphi})\delta\varphi) ~,
\end{equation}
\begin{equation} \label{LCPE2}
\ddot{\Phi}+3\mathbb{H}\dot{\Phi}+[2\dot{\mathbb{H}}-\dot{\bar{k}}+2\mathbb{H}^{2}-\bar{p}\frac{\partial}{\partial \bar{p}}(\frac{\sin \bar{\mu}\gamma \bar{k}}{\bar{\mu}\gamma})^{2}]\Phi=4\pi G(\dot{\bar{\varphi}}\dot{\delta\varphi}-\bar{p}V_{,\bar{\varphi}}(\bar{\varphi})\delta\varphi) ~,
\end{equation}
\begin{equation} \label{LCPE3}
\partial_{a}(\dot{\Phi}+\mathbb{H}\Phi)=4\pi G\dot{\bar{\varphi}}\delta\varphi_{,a}~.
\end{equation}

As we have emphasized in the introduction, among the three classical perturbations
equations (\ref{CPE1}), (\ref{CPE2}) and (\ref{CPE3}) only two are independent.
However, when the gauge-fixing has been done before deriving equations of
motion in the presence of quantum corrections, it may not produce all terms
correctly such that this consistency can not be maintained. In order to preserve the consistency for quantum
corrected equations (\ref{LCPE1}), (\ref{LCPE2}) and (\ref{LCPE3}), the additional
correction terms must be required. The simplest way is only to modify the equations (\ref{LCPE2})
as follow by introducing some additional correction terms,
\begin{equation} \label{LCPE2-1}
\ddot{\Phi}+\{3\mathbb{H}+\frac{1}{\mathbb{H}}[\dot{\mathbb{H}}-\dot{\bar{k}}-\bar{p}\frac{\partial}{\partial \bar{p}}(\frac{\sin \bar{\mu}\gamma \bar{k}}{\bar{\mu}\gamma})^{2}]\}\dot{\Phi}+[2\mathbb{H}^{2}+4\dot{\mathbb{H}}-3\dot{\bar{k}}-3\bar{p}\frac{\partial}{\partial \bar{p}}(\frac{\sin \bar{\mu}\gamma \bar{k}}{\bar{\mu}\gamma})^{2}]\Phi=4\pi G(\dot{\bar{\varphi}}\dot{\delta\varphi}-\bar{p}V_{,\bar{\varphi}}(\bar{\varphi})\delta\varphi) ~.
\end{equation}

The proof of consistency of these equations (\ref{LCPE1}), (\ref{LCPE2-1}) and (\ref{LCPE3}) has been given in the appendix.
Obviously, in the classical limit, $\bar{\mu}\rightarrow 0$,
the equations (\ref{LCPE1}), (\ref{LCPE2-1}) and (\ref{LCPE3}) reduce to the classical cosmological
perturbations (\ref{CPE1}), (\ref{CPE2}) and (\ref{CPE3})
respectively. In addition, we must also point out that since the gauge-fixing has been done before deriving equations of
motion, the introduce of the additional correction terms is not unique. In order to obtain the more complete and unambiguous quantum corrected cosmological perturbations equation, we must consider the gauge invariant variables and derive these perturbations equations in a
gauge invariant manner, which is under progress. Combing these equations, one can obtain the
following second order differential equation for $\Phi$
\begin{eqnarray} \label{LCPE4}
\ddot{\Phi}&-&\nabla^{2}\Phi+\{6\mathbb{H}+2\bar{p}\frac{V_{,\bar{\varphi}}(\bar{\varphi})}{\dot{\bar{\varphi}}}+\frac{1}{\mathbb{H}}[\dot{\mathbb{H}}-\dot{\bar{k}}-\bar{p}\frac{\partial}{\partial \bar{p}}(\frac{\sin \bar{\mu}\gamma \bar{k}}{\bar{\mu}\gamma})^{2}]\}\dot{\Phi}
\nonumber\\
&& \quad
+[8\mathbb{H}^{2}+4\dot{\mathbb{H}}-2\dot{\bar{k}}-4(\frac{\sin \bar{\mu}\gamma \bar{k}}{\bar{\mu}\gamma})^{2}-2\bar{p}\frac{\partial}{\partial \bar{p}}(\frac{\sin \bar{\mu}\gamma \bar{k}}{\bar{\mu}\gamma})^{2}+2\bar{p}\frac{V_{,\bar{\varphi}}(\bar{\varphi})}{\dot{\bar{\varphi}}}\mathbb{H}]\Phi=0~.
\end{eqnarray}

In addition, using the relation between the extrinsic curvature
$\bar{k}$ and the conformal Hubble parameter $\mathbb{H}$
(\ref{RECandHP}), one can obtain
\begin{equation} \label{IextrisicH}
(\frac{\sin \bar{\mu}\gamma \bar{k}}{\bar{\mu}\gamma})^{2}=\frac{1\pm\sqrt{1-4(\bar{\mu}\gamma)^{2}\mathbb{H}^{2}}}{2(\bar{\mu}\gamma)^{2}}~.
\end{equation}

If we denote $S_{h1}=\mathbb{H}^{2}-(\frac{\sin \bar{\mu}\gamma \bar{k}}{\bar{\mu}\gamma})^{2}=\mathbb{H}^{2}-\frac{1\pm\sqrt{1-4(\bar{\mu}\gamma)^{2}\mathbb{H}^{2}}}{2(\bar{\mu}\gamma)^{2}}$, which results from the holonomy corrections in the presence of the metric perturbation, and $S_{h2}=\dot{\mathbb{H}}-\dot{\bar{k}}-\bar{p}\frac{\partial}{\partial \bar{p}}(\frac{\sin \bar{\mu}\gamma \bar{k}}{\bar{\mu}\gamma})^{2}$, which be introduced by the requirement of the consistency, then the above second order differential equation
can be further rewritten as
\begin{equation} \label{LCPE5}
\ddot{\Phi}-\nabla^{2}\Phi+[6\mathbb{H}+2\bar{p}\frac{V_{,\bar{\varphi}}(\bar{\varphi})}{\dot{\bar{\varphi}}}+\frac{S_{h2}}{\mathbb{H}}]\dot{\Phi}
+2[2\mathbb{H}^{2}+\dot{\mathbb{H}}+\bar{p}\frac{V_{,\bar{\varphi}}(\bar{\varphi})}{\dot{\bar{\varphi}}}\mathbb{H}-S_{h2}+2S_{h1}]\Phi=0~.
\end{equation}

Up to now, we have completed the derivation of the cosmological
perturbation equations in the effective loop quantum cosmology
with holonomy corrections.

\section{Discussion}

The effects of quantum gravity on structure formation, generally
called trans-Planckian issues, have been investigated intensively
(for example, we can refer to \cite{trans-Planckian}). In loop
quantum cosmology, the analogous issues have also been
investigated in
Ref.\cite{SuperinflationCFT,PDPLQC,ISPLQC,ConstraintsLQC,XZandYL,MJL}.
However, in Ref.\cite{XZandYL}, they assume that after a
super-inflation phase, the universe underwent a normal inflation
stage. Then they find that the loop quantum effects can hardly
lead to any imprint in the primordial power spectrum. Although in
Ref.\cite{SuperinflationCFT} the scale invariant spectrum was
obtained and the holonomy effects also leave their imprint on the
power spectrum, only the holonomy effects from a fixed background
were taken into account. In Ref.\cite{MJL}, the cosmological
perturbation equations with holonomy corrections were derived
in longitudinal gauge. But they consider only the cases of large scale metric perturbations.
In this paper, along the Hamiltonian
approach we have derived the cosmological perturbation equation
for scalar modes in longitudinal gauge in the presence of holonomy
corrections. In the presence of metric perturbation, we find that
holonomy effects influence both background and perturbations,
which contribute the non-trivial terms $S_{h1}$ and $S_{h2}$. Therefore, the
holonomy effects will affect the power spectrum such that it is
possible that the quantum gravity effects will leave their imprint
on the cosmic microwave background observed today. In the future
work, we will investigate analytically and numerically the
characters of power spectrum in the presence of holonomy
corrections, which might open a window to test the loop quantum
gravity effects.

In addition, when ignoring the additional corrections term $S_{h2}$, which be
introduced by the requirement of the consistency, the second order differential equation
(\ref{LCPE5}) become
\begin{equation} \label{LCPE6}
\ddot{\Phi}-\nabla^{2}\Phi+2(\mathbb{H}-\frac{\ddot{\bar{\varphi}}}{\dot{\bar{\varphi}}})\dot{\Phi}
+2[\dot{\mathbb{H}}-\mathbb{H}\frac{\ddot{\bar{\varphi}}}{\dot{\bar{\varphi}}}+2\mathbb{H}^{2}
+2S_{h1}]\Phi=0~.
\end{equation}
here we have used the Klein-Gordon equation (\ref{KG}).

In this case, we can furthermore introduce the
Mukhanov-Sasaki variable
$\upsilon=\frac{a}{\dot{\bar{\varphi}}}\Phi$. Then the
cosmological perturbation equation (\ref{LCPE6}) reduces to
\begin{equation} \label{LCPEMS}
\ddot{\upsilon}-\nabla^{2}\upsilon
+[(\frac{\ddot{\bar{\varphi}}}{\dot{\bar{\varphi}}})^{.}-(\frac{\ddot{\bar{\varphi}}}{\dot{\bar{\varphi}}})^{2}
+\dot{\mathbb{H}}-\mathbb{H}^{2}+4S_{h1}]\upsilon=0~,
\end{equation}

In momentum space, the cosmological perturbation
equation (\ref{LCPEMS}) can be written as
\begin{equation} \label{LCPEMSM}
\ddot{\upsilon}-[\kappa^{2}-4S_{h1}-m^{2}_{eff}]\upsilon=0~,
\end{equation}
where $\kappa$ denotes the momentum and
$m^{2}_{eff}=(\frac{\ddot{\bar{\varphi}}}{\dot{\bar{\varphi}}})^{.}-(\frac{\ddot{\bar{\varphi}}}{\dot{\bar{\varphi}}})^{2}+\dot{\mathbb{H}}-\mathbb{H}^{2}$.
Therefore, the cosmological perturbation equation (\ref{LCPEMSM})
can be effectively viewed as imposing such a modified dispersion
relation at quantum gravity phenomenological level. Obviously, in
such a modified dispersion relation, both the energy and momentum are
bounded. Here, we point out that, in Ref. \cite{MDRLLW}, Y.
Ling $et.$ $al$ have also proposed a bounded modified dispersion
relation, motivated by the isotropic homogenous effective loop
quantum cosmology with holonomy corrections. Although both are
bounded, they are also very different, implying we can not simply
use the background corrections instead of perturbation
corrections. In the future work, we will furthermore discuss the
implications of such two modified dispersion relations.

Our present paper is the first step towards studying the holonomy
corrected cosmological perturbation equations in the presence of
metric perturbation. Since constraints are modified, the form of
gauge invariant variables should change as well. Therefore, it is
necessary to study the perturbations with different gauges or in a
gauge invariant manner in this formalism, which is under progress.

\section*{Acknowledgement}

J. P. Wu is grateful to Prof. Yongge Ma and Wei-Jia Li for helpful
discussion. J. P. Wu is partly supported by NSFC(No.10975017). Y. Ling
is partly supported by NSFC(No.10875057), Fok Ying Tung Education
Foundation(No. 111008), the key project of Chinese Ministry of
Education(No.208072) and Jiangxi young scientists(JingGang Star)
program. He also acknowledges the support by the Program for
Innovative Research Team of Nanchang University.

\begin{appendix}
\section{The proof of consistency of these equations (\ref{LCPE1}), (\ref{LCPE2-1}) and (\ref{LCPE3})}

In this appendix, we will give a proof of the consistency of these equations (\ref{LCPE1}), (\ref{LCPE2-1}) and (\ref{LCPE3}).
Without loss of generality, we will only derive the Eq. (\ref{LCPE2-1}) from the Eqs. (\ref{LCPE1}) and (\ref{LCPE3}).
From the perturbation equation (\ref{LCPE3}), by taking the spatial derivation, we can obtain
\begin{equation} \label{LCPE3-1}
\frac{d}{d\eta}(\nabla^{2}\Phi)+\mathbb{H}\nabla^{2}\Phi-4\pi G\dot{\bar{\varphi}}\nabla^{2}\delta\varphi=0~.
\end{equation}

In addition, using the corrected Raychaudhuri equation (\ref{CRayE}) and the perturbation equation (\ref{LCPE1}),
the term $\nabla^{2}\Phi$ can be expressed as
\begin{equation} \label{LCPE3-1}
\nabla^{2}\Phi=3\mathbb{H}\dot{\Phi}
+\left[6\mathbb{H}-8\pi G(\dot{\bar{\varphi}}^{2}+\bar{p}V(\bar{\varphi}))\right]\Phi
+4\pi G(\dot{\bar{\varphi}}\dot{\delta \varphi}+\bar{p}V_{,\bar{\varphi}}(\bar{\varphi}) \delta \varphi)~.
\end{equation}

Therefore, we can obtain the following expressions:
\begin{eqnarray} \label{a}
\frac{d}{d\eta}(\nabla^{2}\Phi)
&=& 3 \mathbb{H} \ddot{\Phi} + [3\dot{\mathbb{H}} + 6 \mathbb{H}^{2} - 8 \pi G ( \dot{\bar{\varphi}}^{2} + \bar{p} V(\bar{\varphi}) ) ]\dot{\Phi}
\nonumber\\
&& \quad
+[12\mathbb{H}\dot{\mathbb{H}}-8 \pi G( 2\dot{\bar{\varphi}} \ddot{\bar{\varphi}} + \dot{\bar{p}} V(\bar{\varphi}) + \bar{p} \dot{\bar{\varphi}} V_{,\bar{\varphi}}(\bar{\varphi}) ) ] \Phi
\nonumber\\
&& \quad
+4\pi G [\ddot{\bar{\varphi}} \dot{\delta \varphi} + \dot{\bar{\varphi}} \ddot{\delta \varphi} + \dot{\bar{p}} V_{,\bar{\varphi}}(\bar{\varphi}) \delta \varphi
+ \bar{p} V_{,\bar{\varphi}}(\bar{\varphi}) \dot{\delta \varphi} + \bar{p} \dot{\bar{\varphi}} V_{,\bar{\varphi}\bar{\varphi}}(\bar{\varphi}) \delta \varphi]~,
\end{eqnarray}
\begin{eqnarray} \label{b}
\mathbb{H}\nabla^{2}\Phi
= 3\mathbb{H}^{2}\dot{\Phi} + [6 \mathbb{H}^{3} - 8 \pi G \mathbb{H}( \dot{\bar{\varphi}}^{2} + \bar{p} V(\bar{\varphi}))]\Phi
+ 4 \pi G \mathbb{H} [ \dot{\bar{\varphi}} \dot{\bar{\varphi}} + \bar{p} V_{,\bar{\varphi}}(\bar{\varphi}) \delta \varphi]~.
\end{eqnarray}

In addition, we can also expressed the term $4 \pi G \dot{\bar{\varphi}} \nabla^{2} \delta \varphi$ as following with the help of (\ref{PKG})
\begin{eqnarray} \label{c}
4 \pi G \dot{\bar{\varphi}} \nabla^{2} \delta \varphi = 4 \pi G \dot{\bar{\varphi}} [\delta\ddot{\varphi} + 2 \mathbb{H} \dot{\bar{\varphi}} + \bar{p} V_{,\bar{\varphi}\bar{\varphi}}(\bar{\varphi}) \delta\varphi
+ 2 \bar{p} V_{,\bar{\varphi}}(\bar{\varphi}) \Phi - 4 \dot{\bar{\varphi}} \dot{\Phi} ]~.
\end{eqnarray}

Collecting all the above expressions (\ref{a}), (\ref{b}) and (\ref{c}), we can obtain the perturbation equation (\ref{LCPE2-1}) by straightly calculating.
Similarly, we can also derive equation (\ref{LCPE1}) or (\ref{LCPE3}) from the remaining two equations. Therefore, among these equations above only
two of them are independent.

\end{appendix}


\begin{thebibliography}{99}


\bibitem{ReviewLQC}
M. Bojowald, Loop quantum cosmology, Living Rev. Relativity 8, 11, (2005) [gr-qc/0601085].


\bibitem{AvoidSingularity}
M. Bojowald, Absence of Singularity in Loop Quantum Cosmology, Phys. Rev. Lett. 86 (2001) 5227 [gr-qc/0102069].

\bibitem{QNBBofATP}
A. Ashtekar, T. Pawlowski, and P. Singh, Quantum nature of the big bang: an analytical and numerical investigation, Phys. Rev. D 73:124038, (2006) [gr-qc/0604013].

\bibitem{QNBBIMofATP}
A. Ashtekar, T. Pawlowski, and P. Singh, Quantum nature of the big bang: improved dynamics, Phys.Rev.D74:084003,2006,  [gr-qc/0607039].

\bibitem{AISS}

A. Ashtekar, M. Bojowald and J. Lewandowski, Mathematical structure of loop quantum cosmology, Adv. Theor. Math. Phys. 7, 233 (2003) [gr-qc/0304074];

M. Bojowald, G. Date and K. Vandersloot, Homogeneous loop quantum cosmology: The role of the spin connection, Class. Quant. Grav. 21, 1253 (2004) [gr-qc/0311004];

P. Singh and A. Toporensky, Big crunch avoidance in k = 1 loop quantum cosmology, Phys. Rev. D 69, 104008 (2004) [gr-qc/0312110];

G. V. Vereshchagin, Qualitative approach to semi-classical loop quantum cosmology, JCAP 0407, 013 (2004) [gr-qc/0406108];

G. Date, Absence of the Kasner singularity in the effective dynamics from loop quantum cosmology, Phys. Rev. D 71, 127502 (2005) [gr-qc/0505002];

G. Date, Absence of the Kasner singularity in the effective dynamics from loop quantum cosmology, Phys. Rev. D 71, 127502 (2005) [gr-qc/0505002];

G. Date and G. M. Hossain, Genericity of big bounce in isotropic loop quantum cosmology, Phys. Rev. Lett. 94, 011302 (2005) [gr-qc/0407074];

R. Goswami, P. S. Joshi and P. Singh, Quantum evaporation of a naked singularity, Phys. Rev. Lett. 96, 031302 (2006) [gr-qc/0506129].


\bibitem{BouncingofLQCB}
M. Bojowald, The Early Universe in Loop Quantum Cosmology, J. Phys. Conf. Ser. 24 (2005) 77, [gr-qc/0503020].

\bibitem{BouncingofLQCTM}
T. Stachowiak and M. Szydlowski, Exact solutions in bouncing cosmology, Phys.Lett.B646:209-214,2007, Phys. Lett. B 646 (2007) 209 [gr-qc/0610121].


\bibitem{Superinflation}
M. Bojowald, Inflation from quantum geometry, Phys. Rev. Lett. 89 (2002) 261301 [gr-qc/0206054].

\bibitem{SuperinflationCFT}
E. J. Copeland, D. J. Mulryne, N. J. Nunes, and M. Shaeri, Super-inflation in loop quantum cosmology, Phys.Rev.D77:023510,2008, [arXiv:0708.1261].


\bibitem{HamPerturb}
M.\ Bojowald, H.\ Hern\'andez, M.\ Kagan, P.\ Singh, and A.\ Skirzewski, Hamiltonian cosmological perturbation theory with loop quantum gravity corrections, {\em Phys.\ Rev.\ D} 74 (2006) 123512, [gr-qc/0609057]

\bibitem{FormationBPRL}
M.\ Bojowald, M.\ Kagan, P.\ Singh, and A.\ Skirzewski, H.\ Hern\'andez, Formation and evolution of structure in loop cosmology, {\em Phys.\ Rev.\ D} 74 (2006) 123512, [astro-ph/0611685].


\bibitem{GaugeInvariant}
M.\ Bojowald, G. M. Hossain, M.\ Kagan, and S.\ Shankaranarayanan, Gauge invariant cosmological perturbation equations with corrections from loop quantum gravity, Phys.Rev.D79:043505,2009, [arXiv:0811.1572].

\bibitem{AnomalyFreedom}
M.\ Bojowald, G. M. Hossain, M.\ Kagan, and S.\ Shankaranarayanan, Anomaly freedom in perturbative loop quantum gravity, Phys. Rev. D 78 (2008) 063547, [arXiv:0806.3929].



\bibitem{VectorB}
M.\ Bojowald, G. M. Hossain, Cosmological vector modes and quantum gravity effects, Class.Quant.Grav.24:4801-4816,2007, [arXiv:0709.0872].

\bibitem{TensorB}
M.\ Bojowald, G. M. Hossain, Loop quantum gravity corrections to gravitational wave dispersion, Phys.Rev.D77:023508,2008, [arXiv:0709.2365].




\bibitem{PDPLQC}
G. M. Hossain, Primordial Density Perturbation in Effective Loop Quantum Cosmology, Class.Quant.Grav. 22 (2005) 2511, [gr-qc/0411012].

\bibitem{ISPLQC}
G. Calcagni, M. Cortes, Inflationary scalar spectrum in loop quantum cosmology, Class.Quant.Grav. 24 (2007) 829-854, [gr-qc/0607059].

\bibitem{ConstraintsLQC}
D. J. Mulryne, N. J. Nunes, Constraints on a scale invariant power spectrum from superinflation in LQC, Phys.Rev. D74 (2006) 083507, [astro-ph/0607037].


\bibitem{XZandYL}
X. Zhang, Y. Ling, Inflationary universe in loop quantum cosmology, JCAP 0708:012,2007, [arXiv:0705.2656].

\bibitem{MJL}
M. Artymowski, Z. Lalaky and  L. Szulcz, Loop Quantum Cosmology corrections to in
ationary models, JCAP 0901:004,2009, [arXiv:0807.0160].


\bibitem{Tsujikawa}
S. Tsujikawa, P. Singh, and R. Maartens, Loop quantum gravity effects on inflation and the CMB, Class.Quant.Grav. 21 (2004) 5767-5775, [astro-ph/0311015].


\bibitem{Shimano}
M. Shimano, T. Harada, Observational constraints of a power spectrum from super-inflation in Loop Quantum Cosmology, Phys.Rev.D80:063538,2009, [arXiv:0909.0334].






\bibitem{GWLQC}

J. Mielczarek, M. Szydlowski, gravitons as the observable for Loop Quantum Cosmology, Phys.Lett.B657:20-26,2007, [arXiv:0705.4449];

J. Mielczarek, M. Szydlowski, Relic gravitons from super-inflation, [arXiv:0710.2742];

J. Mielczarek, Gravitational waves from the Big Bounce, JCAP0811:011,2008, [arXiv:0807.0712];

J. Grain, A. Barrau, Cosmological footprints of loop quantum gravity, Phys. Rev. Lett. 102, 081301 (2009), [arXiv:0902.0145];

J. Mielczarek, Tensor power spectrum with holonomy corrections in LQC, Phys.Rev.D79:123520,2009, [arXiv:0902.2490];

J. Grain, T. Cailleteau, A. Barrau, and A. Goreck, Fully LQC-corrected propagation of gravitational waves during slow-roll inflation, [arXiv:0910.2892];

J. Grain, Loop Quantum Cosmology corrections on gravity waves produced during primordial inflation, published in the AIP Proceedings of the 'Invisible Universe International Conference', UNESCO-Paris, June 29-July 3, 2009, [arXiv:0911.1625].




\bibitem{CosmologicalPerturbations}
V.F. Mukhanov, H.A. Feldman and R.H. Brandenberger, Theory of cosmological perturbations, Phys. Rept. 215, 203 (1992).



\bibitem{muScheme}
A. Corichi and P. Singh, Is loop quantization in cosmology unique? Phys.Rev.D78:024034,2008, [arXiv:0805.0136].


\bibitem{trans-Planckian}
J. Martin, R. H. Brandenberger, The Trans-Planckian Problem of Inflationary Cosmology, Phys.Rev.D63:123501,2001, [hep-th/0005209];

R. H. Brandenberger, Trans-Planckian physics and inflationary cosmology, [hep-th/0210186].




\bibitem{MDRLLW}
Y. Ling, W. J. Li, and J. P. Wu, Bouncing universe from a modified dispersion relation, JCAP 11 (2009) 016, [arXiv:0909.4862].





\end{thebibliography}
\end{document}